\documentclass[pra,twocolumn,showpacs,amsmath,amssymb,superscriptaddress,unsortedaddress,floatfix,longbibliography]{revtex4-1}
\usepackage{graphicx}
\usepackage{latexsym}
\usepackage{amsmath}
\usepackage{graphics}
\usepackage{amssymb}
\usepackage{layout}
\usepackage{verbatim}
\usepackage{amsfonts,epsfig,dsfont}
\usepackage{natbib}
\usepackage{hyperref}
\usepackage{color}
\usepackage[latin1]{inputenc}       
\usepackage[T1]{fontenc}
\usepackage[outdir=./]{epstopdf}

\newcommand{\ket}[2][]{{|#2\rangle_{#1}}}
\newcommand{\bra}[2][]{{}_{#1}\langle #2|}

\newcommand{\tr}{\textrm{Tr}}

\def\duzomniejsze{<\kern-.7mm<}
\def\duzowieksze{>\kern-.7mm>}

\def\textbf#1{{\bf #1}}
\def\beq{\begin{equation}}
\def\eeq{\end{equation}}
\def\be{\begin{equation}}
\def\ee{\end{equation}}
\def\ben{\begin{eqnarray}}
\def\een{\end{eqnarray}}
\def\beqa{\begin{eqnarray}}
\def\eeqa{\end{eqnarray}}
\def\eea{\end{array}}
\def\bea{\begin{array}}
	\newcommand{\unit}{\mathbb{I}}

	\usepackage{color}
	
	\usepackage{ulem} 
	
\begin{document}
	
	\title{Qubit-environment Negativity versus Fidelity of conditional environmental
	states for an NV-center spin qubit interacting with a nuclear environment} 
	
	\author{Ma{\l}gorzata Strza{\l}ka}
	\affiliation{Department of Theoretical Physics, Wroc{\l}aw University of Technology,
		50-370 Wroc{\l}aw, Poland}
	
	\author{Damian Kwiatkowski}
	\affiliation{Institute of Physics, Polish Academy of Sciences, 02-668 Warsaw, Poland}
	
	\author{{\L}ukasz Cywi{\'n}ski}
	\affiliation{Institute of Physics, Polish Academy of Sciences, 02-668 Warsaw, Poland}
	
	\author{Katarzyna Roszak}
	\affiliation{Department of Theoretical Physics, Wroc{\l}aw University of Technology,
		50-370 Wroc{\l}aw, Poland}
	
	\date{\today}

	\begin{abstract}
		We study the evolution of qubit-environment entanglement, quantified using Negativity, 
		for NV-center spin qubits 
		interacting with an environment of $^{13}$C isotope partially polarized nuclear spins in the diamond lattice.
		We compare it with the evolution of the Fidelity of environmental states 
		conditional on the pointer states of the qubit, which can serve as a tool to distinguish
		between entangling and non-entangling decoherence in pure-dephasing scenarios.  
		The two quantities show remarkable agreement during the evolution
		in a wide range of system parameters, leading to the conclusion
		that the amount of entanglement generated between the qubit and the
		environment is proportional to the trace that the joint evolution leaves in the environment.
	\end{abstract}

	\maketitle

	\section{Introduction \label{sec1}}
	
	The study of system-environment or even qubit-environment entanglement
	is seriously limited due to large sizes of the studied environments, which 
	translates into few entanglement measures being available on the level
	of density matrix considerations. 
	This would not be a problem if the joint system-environment state would
	be pure, but in most realistic scenarios the initial state of the environment
	is far from pure except for extremely low temperatures. 
	In fact, the only measure which can serve
	to quantify entanglement between two systems of any size which can be calculated
	directly from the joint density matrix is Negativity \cite{vidal02,lee00a} (or closely related
	logarithmic Negativity \cite{plenio05b}), which nevertheless requires diagonalization
	of a matrix of the same size as the joint systems' Hilbert space.
	Negativity has its limitations, since there exist entangled states 
	which are not detected by it \cite{Horodecki_PLA97,Horodecki_PRL98}, but it is the best available tool
	if the two potentially entangled systems are larger than a qubit and a qutrit
	and the purity of the system is less than one.
	All other measures require some form of minimization over possible 
	representations of the states in different bases, which becomes
	highly cumbersome with growing system size \cite{Mintert_PR05,Plenio_QIC07,Horodecki_RMP09,Aolita_RPP15,Kraus_PRA00}.
	
	Recently, relatively straightforward methods for detecting system-environment \cite{roszak18}
	and qubit-environment \cite{roszak15a} entanglement have been found
	for a scenario limiting the type of Hamiltonians that drive the
	system-environment interaction. The type of Hamiltonians under study
	leads to pure dephasing of the system/qubit when the 
	environment is traced out (it cannot cause transitions between 
	a set of system pointer states \cite{Zurek_PRD81,Zurek_RMP03} which is singled out by the interaction
	itself).	
	This may seem like a large loss of generality, but said type of 
	system-environment interactions are abundant, especially in solid state
	scenarios \cite{Medford_PRL12,Kawakami_PNAS16,borri01,vagov04,roszak06a,Paladino_RMP14,Szankowski_JPCM17,Witzel_PRB06,Yao_PRB06,Cywinski_PRB09,Coish_PRB10,Bluhm_NP10,Yang_RPP17,Malinowski_NN17,Kwiatkowski_PRB18},
	but also for trapped ions \cite{Biercuk_Nature09,Monz_PRL11}.
	
	The problem of the method is that it does not quantify the amount of entanglement,
	instead answering the question if system-environment entanglement is
	present at a given time after initialization of the system/qubit in a pure state.
	As there are no limitations on the initial state of the environment, which is
	likely to be mixed, the whole system is initially impure, and pure dephasing can occur
	either while being accompanied by entanglement generation or due to completely separable
	system-environemnt evolutions \cite{Eisert_PRL02,Hilt_PRA09,Pernice_PRA11,roszak15a,roszak18}.
	This is in stark contrast to pure state system-environment evolutions, for which
	pure dephasing is irrefutably linked with the buildup of entanglement with
	the environment \cite{Zurek_RMP03,Hornberger_LNP09}.
	The results of Refs \cite{roszak15a,roszak18} show that system-environment entanglement leaves a detectable trace 
	on the environment, while it is impossible to determine entangling from non-entangling
	evolutions by straightforward measurements of system pure dephasing. 
	More involved schemes for the detection of qubit-environment entanglement
	by operations and measurements on only the qubit subsystem have been
	recently proposed \cite{roszak19}. Both, detection of entanglement by measurement 
	only on the environment and 
	detection of entanglement via operations on the qubit are possible 
	because the problem is restricted to a special class of Hamiltonians, hence there is no contradiction with the popular theorem on the impossibility of local detection of entanglement.
	
	The evolution of a qubit and its environment is not accompanied by entanglement
	generation if and only if the evolution of the states of the environment 
	conditional on the pointer states of the qubit is the same at all times
	(if this occurs only at isolated points of time, then there is no entanglement
	only at these times) \cite{roszak15a},
	\begin{equation}
	\label{war0}
	\hat{R}_{00}(t)=\hat{R}_{11}(t),
	\end{equation}
	where said conditional states are denoted by $\hat{R}_{ii}(t)$, with $i=0,1$,
	and they correspond to the state the environment would be in at time $t$ if 
	the qubit were initialized in pointer state $|i\rangle$ at the initial time.
	Hence if the qubit is initialized in a superposition state $a|0\rangle +b|1\rangle$,
	the state of the environment at time $t$, obtained by tracing out the qubit from
	the full qubit-environment density matrix, is given by
	\begin{equation}
	\hat{R}(t)=|a|^2\hat{R}_{00}(t)+|b|^2\hat{R}_{11}(t).
	\end{equation}
	In case there is no qubit-environment entanglement, this state is the same regardless
	on the initial qubit superposition and is equal to the state the environment would
	evolve to under the influence of the qubit in one of its pointer states.
	When there is entanglement generated with the environment, the situation is
	qualitatively different, and the state of the environment depends on the probability
	of finding the qubit in either pointer state.
	Therefore we can talk about a
	trace left by joint qubit-environment evolution on the environment
	which is present only for entangling evolutions.
	
	Here, we make a first step towards a measure of qubit-environment entanglement,
	designed to quantify the amount of entanglement generating during evolutions
	of pure dephasing type. To this end, we test if the magnitude of the trace
	left by entangling evolutions on the state of the environment is proportional
	to the amount of actual entanglement generated on a realistically modeled
	NV-center in diamond spin qubit interacting with a nuclear spin environment 
	 \cite{Kwiatkowski_PRB18}.
	The choice of test system is based both on its experimental relevance \cite{Staudacher_Science13,DeVience_NN15,Haberle_NN15,Lovchinsky_Science16,Wrachtrup_JMR16,Degen_RMP17}
	and on the wide variety of test scenarios it gives. 
	As the NV-center has effectively spin $S=1$, the spin states form a qutrit,
	but the uneven level spacing between the different spin states allows 
	for any two levels out of three to be singled out as the qubit under study.
	Furthermore, this type of spin qubits
	interact strongly only with nuclei of spinful carbon isotopes $^{13}$C, which are few within the
	diamond crystal lattice, and both their number and locations vary, which 
	leads to different evolutions. The whole qubit-environment Hilbert space is 
	therefore small enough to allow for effective diagonalization of 
	matrices within it, which is necessary to find the evolution of Negativity.
	
	We test a number of qubit-environment evolutions
	driven by different interaction Hamiltonians, all within the NV-center spin qubit
	model with five
	relevant environment nuclei. We find a remarkable
	agreement between the time-evolution of the entanglement measure Negativity
	and the Fidelity between the states of the environment conditional on the qubit
	pointer states (the difference between the two conditional environmental states
	serves to test for qubit-environment entanglement in Ref.~\cite{roszak15a}).
	Furthermore we find that this agreement is also present for evolutions 
	which cannot be detected by the qubit-based scheme of Ref.~\cite{roszak19}.
	We conjecture that the effect is of more general nature and that 
	said Fidelity could be the basis of an entanglement measure designed 
	specifically for pure dephasing evolutions.
	
	The paper is organized as follows. We introduce the NV center qubit 
	and its environment in Sec.~\ref{sec2}. In Sec.~\ref{sec3}
	we provide the definitions necessary to calculate qubit-environment
	Negativity. In Sec.~\ref{sec4} we discuss the correlation between entanglement
	generation in pure dephasing scenarios and the difference between conditional
	evolution of the environment and use the Fidelity to quantify the difference.
	Results obtained for realistically modeled spin qubits with randomly chosen 
	environments are presented and discussed in Sec.~\ref{sec5}, while Sec.~\ref{sec6}
	contains concluding remarks.

	\section{NV center interacting with a partially polarized
		nuclear environment \label{sec2}}
	
	Our test system consists of a spin qubit defined on an NV center in diamond
	interacting with an environment of nuclear spins of the spinful carbon
	isotope, $^{13}$C. As most of the diamond crystal lattice consists of spinless 
	carbon nuclei, the relevant atoms of the environment (for decoherence)
	are few and randomly located. This is of use for testing of the correlation
	between generated entanglement and the magnitude of the trace that entangling
	evolutions leave on the conditional states of the environment
	(how strongly the conditional states of the environment are affected by
	entangling evolution), since 
	the resulting system-environment evolutions vary
	depending on the choice of qubit as well as depending on the locations
	of the relevant carbon isotope atoms and their number.
	
	The low energy states of the center constitute an effective electronic spin $S = 1$,
	so we are dealing with a qutrit defined on the $m = -1$, $0$ and $1$ levels,
	subsequently labeled as $|-1\rangle$, $|0\rangle$ and $|1\rangle$.
	 This is
	 subjected to a zero-field splitting $\Delta (S^{z})^2$, with the direction of $z$ axis determined by the geometry of the center, so the presence of a magnetic field along the $z$ axis leads to a splitting of the $m_s=\pm 1$ levels
	 and an uneven level spacing between them. 
	 This allows for any two-level subspace to be used as a qubit controlled by microwave electromagnetic fields. We choose 
	 two out of the three possible qubits for our study, one is
	 the most widely employed qubit based on the $m = 0$ and $1$ levels, and the other is based on 
	 the $m = -1$ and $1$ levels.
	
The large value of the zero-field splitting, $\Delta=2.87$ GHz and a large ratio of electronic and nuclear gyromagnetic factors
lead to the suppression of transitions between the qutrit states mediated by the environment, hence the system can be described as one which undergoes
only pure dephasing type of interaction \cite{Zhao_PRB12}. 
Additionally, the $|0\rangle$ state is decoupled from the environment,
so the qutrit-environment Hamiltonian is of the form
	\begin{eqnarray}
	\nonumber
	\hat{H} &=& (\Delta +\gamma_{e}B_{z}) \ket{-1}\bra{-1} + (\Delta -\gamma_{e}B_{z}) \ket{1}\bra{1}+\hat{H}_E\\
	\label{ham}
	&& -
	\ket{-1}\bra{-1} \otimes\hat{V} +
	\ket{1}\bra{1} \otimes\hat{V}.
	\end{eqnarray}
	The first two terms in the Hamiltonian describe the free evolution of the qutrit.
	The energy of states $|\pm 1\rangle$ depend on the zero-field splitting symmetrically 
	and asymmetrically on a magnetic-field-dependent term, where $\gamma_e=28.08$ MHz/T
	is the electron gyromagnetic ratio.
	This part of the Hamiltonian commutes with all other terms in eq.~(\ref{ham})
	and the resulting evolution can therefore be eliminated from the joint
	system-environment evolution via a unitary operation performed solely on the qutrit
	(by moving to a rotation frame with respect to the qubit).
	The consequence of this is that it has no bearing on either the generation
	of entanglement or on its magnitude.
	It will also play no part in the conditional evolution of the environment.
	
	The second term in the Hamiltonian describes the free evolution of environmental
	spins,
	\begin{equation}
	\label{he}
	\hat{H}_E=\sum_{k}\gamma_{n}B_{z}\hat{I}^{z}_{k},
	\end{equation}
	where $k$ labels the spins, $\gamma_{n} \! = \! 10.71$ MHz/T
	is the gyromagnetic ratio for $^{13}$C nuclei, $\hat{I}^{z}_{k}$ is the operator of the $z$ component of nuclear spin $k$. A term describing the internuclear magnetic dipolar interactions has been omitted, since the free evolution decoherence process occurs on much shorter timescales than said interactions (in contrast to coherence observed in spin echo experiment \cite{Zhao_PRB12,Kwiatkowski_PRB18}).
	
	The last term in eq.~(\ref{ham}) describes the hyperfine interaction
	between the spin qubit and its nuclear spin environment.
	It is given by
	\beq
	\label{v}
	\hat{V} = \sum_{k}\sum_{j \in (x,y,z)} \mathbb{A}^{z,j}_k\hat{I}^{j}_{k} \,\, .
	\eeq
	If we omit the Fermi contact interaction \cite{gali08}
	which is related to the non-zero probability of finding an electron bound to the NV center on the location of a given nucleus, and only take the dipolar coupling
	into account, the coupling constants present in eq.~(\ref{v})
	are given by
	\begin{equation}
	\label{a}
	\mathbb{A}^{z,j}_k=\frac{\mu_0}{4\pi}\frac{\gamma_e\gamma_n}{r_{k}^3}\left(1-\frac{3(\mathbf{r}_{k}\cdot\hat{\mathbf{j}})(\mathbf{r}_{k}\cdot\hat{\mathbf{z}})}{r_{k}^2}\right).
	\end{equation}
	Here, $\mu_0$ is the magnetic permeability of the vacuum, $\mathbf{r}_{k}$ is a displacement vector between the $k$-th nucleus and the qubit and
	$\hat{\mathbf{j}}=\hat{\mathbf{x}},\hat{\mathbf{y}},\hat{\mathbf{z}}$
	denote versors corresponding to three distinct directions.
	
	Note that the free evolution of the environment and the interaction term
	do not commute for non-zero magnetic fields, therefore the free evolution
	cannot be eliminated via a local unitary transformation and can
	take part in the generation of qubit-environment entanglement, regardless of the qubit of choice.
	Hence, the evolution operator for the qutrit and the environment
	(without the irrelevant free evolution of the qutrit)
	is given by
	\begin{equation}
	\hat{U}(t)=\sum_{m=-1}^1|m\rangle\langle m|\otimes \hat{w}_m(t),
	\end{equation}
	with 
	\begin{subequations}
		\label{w}
		\begin{eqnarray}
		\hat{w}_{-1}(t)&=&e^{-\frac{i}{\hbar}(\hat{H}_E-\hat{V}) t},\\
		\hat{w}_0(t)&=&e^{-\frac{i}{\hbar}\hat{H}_E t},\\
	\hat{w}_1(t)&=&e^{-\frac{i}{\hbar}(\hat{H}_E+\hat{V}) t}.
		\end{eqnarray}
	\end{subequations}
	
	In the following we will be considering an initial state which is a product
	of a pure state of the qutrit within one of the two chosen qubit subspaces, $|\psi\rangle =a|0\rangle+b|1\rangle$ or $|\psi\rangle =a|-1\rangle+b|1\rangle$, and 
	a partially polarized state of the nuclear environment, $\hat{R}(0)$,
	(which is mixed),
	\begin{equation}
	\hat{\sigma}(0)=|\psi\rangle\langle\psi|\otimes\hat{R}(0).
	\end{equation}
	The Hamiltonian (\ref{ham}) does not contain any terms which allow for transitions 
	between different qutrit pointer states $|m\rangle$,
	so the effectively the evolution
	of such an initial state is governed only by the term in the Hamiltonian which 
	contain the relevant qubit states, so either $|0\rangle$ and $|1\rangle$
	or $|-1\rangle$ and $|1\rangle$.
	We assume that $\hat{R}(0)$ does not contain any correlations between the nuclei, 
	so
	$\hat{R}(0) \! = \! \bigotimes_{k} \hat{\rho}_{k}$, where $\hat{\rho}_{k}$ is the density matrix of $k$-th nucleus, given in the case of spin-$1/2$ nuclei by
	\begin{equation}
	 \hat{\rho}_k =\frac{1}{2}(\mathds{1} + 2p_{k}\hat{I}^{z}_{k}),
	\end{equation} 
	where $p_k \! \in \! [-1,1]$ is the polarization of the $k$-th nucleus.  
	Without dynamic nuclear polarization, $p_{k}=0$ for all $k$, the density operator of the environment at low fields is $\hat{R}(0) \! \propto \! \mathds{1}$, and according to the
	results of Ref.~\cite{roszak15a} no qubit environment entanglement would
	form throughout the evolution.
	Since such nuclear polarization of the environment for an NV center has been recently mastered \cite{London2013,Fischer2013a,Pagliero2018,Wunderlich2017,Alvarez2015,King2015,  Scheuer2017,Poggiali2017,Hovav2018}, the assumption of the specially prepared initial state of the
	environment is reasonable. In the following we will assume that the polarizations of each
	environmental nucleus are the same, so $p_k=p$ for all $k$.
	
	Since both the initial state and the evolution operator are known,
	we can write the time-evolved qubit-environment density matrix in the form
	\begin{equation}
	\label{mac1}
	\tilde{\sigma}(t)=\left(
	\begin{array}{cc}
	|a|^2\hat{R}_{nn}(t)&ab^*\hat{R}_{n1}(t)\\
	a^*b\hat{R}_{1n}(t)&|b|^2\hat{R}_{11}(t)
	\end{array}\right),
	\end{equation}
	with $n=-1,0$ depending on the choice of the qubit.
	Here the environmental operators $\hat{R}_{ij}(t)$
	are given by
	\begin{equation}
	\label{rij}
	\hat{R}_{ij}(t)=\hat{w}_i(t)\hat{R}(0)\hat{w}_j^{\dagger}(t).
	\end{equation}
	
	\section{Negativity - an entanglement measure applicative for large
		bipartite systems \label{sec3}}
	
	For large bipartite systems, such as the studied here qubit and environment
	(where the latter is larger), the choice of entanglement measures which
	can be computed is very limited. It comes down in fact practically 
	to the choice between Negavitity \cite{vidal02,lee00a} or logarithmic
	Negativity \cite{plenio05b}. Both measures are closely related and are
	based on the positive partial transpose (PPT) criterion
	of separability \cite{Peres_PRL96,Horodecki_PLA96}.
	The criterion and therefore also the measures do not detect a certain
	type of entangled states called bound entangled states \cite{Horodecki_PLA97,Horodecki_PRL98}, but in the studied scenario, 
	namely in the case of an initially pure-state qubit, bound entanglement
	never forms 
	\cite{horodecki00,roszak15a}.
	Therefore in what follows, Negativity (and logarithmic Negativity)
	signifies separability if and only if the joint qubit and environment
	state is really separable.
	
	In what follows, we choose to employ plain Negativity.
	It is defined as the absolute sum of the negative eigenvalues
	of the density matrix of the whole system after a partial transposition 
	with respect to one of the two potentially entangled subsystems
	and can be written as
	\begin{equation}
	\label{neg}
	N(\hat{\sigma})=\sum_i\frac{|\lambda_i|-\lambda_i}{2},
	\end{equation}
	where $\lambda_i$ denote all eigenvalues of
	the density matrix after partial transposition, $\hat{\sigma}^{\Gamma_A}$.
	Obviously the positive eigenvalues cancel out in eq.~(\ref{neg})
	while only negative eigenvalues are left.
	Negativity does not depend on the system with respect to which 
	partial transposition is performed, $A=Q,E$.
	
	We calculate negativity at each instance of time by first performing
	partial transposition with respect to the qubit
	on the time-evolved qubit-environment density matrix (\ref{mac1}),
	\begin{equation}
	\label{mac2}
	\tilde{\sigma}^{\Gamma_Q}(t)=\left(
	\begin{array}{cc}
	|a|^2\hat{R}_{nn}(t)&a^*b\hat{R}_{1n}(t)\\
	ab^*\hat{R}_{n1}(t)&|b|^2\hat{R}_{11}(t)
	\end{array}\right),
	\end{equation}
	and then finding the eigevalues
	of the matrix obtained in this way.
			
	\section{Fidelity of conditional environmental states \label{sec4}}
	
	As shown in Refs \cite{roszak15a,roszak18}, the if and only if criterion of 
	separability for pure-dephasing qubit-environment evolutions
	at time $t$
	can be written as
	\begin{equation}
	\label{war}
	\hat{R}_{nn}(t)=\hat{R}_{11}(t),
	\end{equation}
	where the density matrices of the environment conditional on the qubit being in either
	of its pointer states is given by eq.~(\ref{rij}).
	This means that there is no entanglement between the qubit and the
	environment at time $t$,
	for an initial state that involves a pure state superposition in the qubit subspace, if and only if the environment would be in the same state at time $t$ if the 
	qubit would have been initialized in either of its pointer states.
	
	\begin{figure}[tb]
		\includegraphics[width=1\columnwidth]{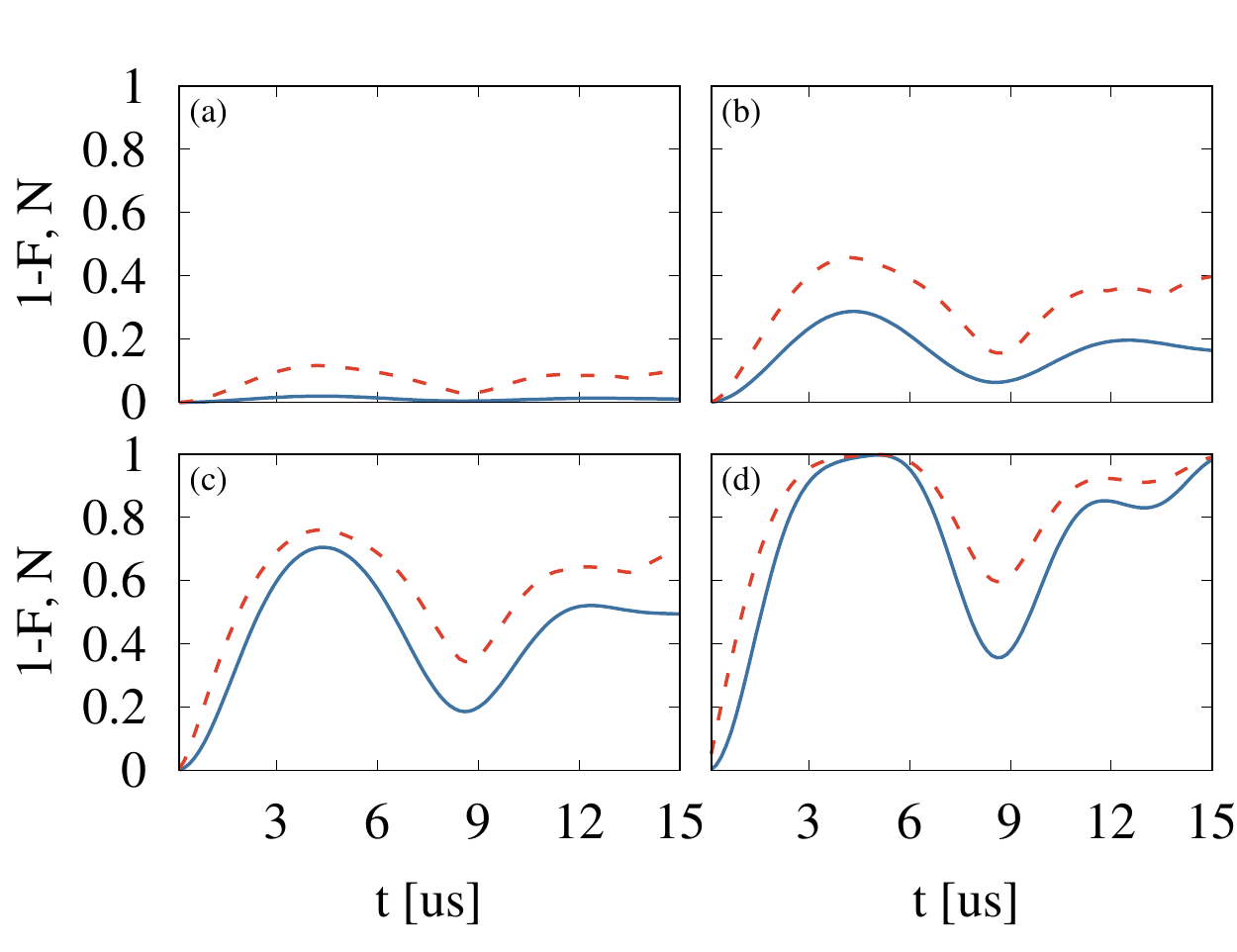}
		\caption{Evolution of qubit-environment Negativity (red dashed lines) and one-minus-Fidelity
			between conditional
			environmental states (blue solid lines) 
			for a qubit defined on $m=0$ and $m=1$ spin states
			and five environmental spins at random locations
			as a function of time for zero magnetic field and different initial polarizations 
			of the environment: (a) $p=0.1$, (b) $p=0.4$, (c) $p=0.7$, (d) $p=1$.  
		}\label{fig1}
	\end{figure}

	\begin{figure}[tb]
		\includegraphics[width=1\columnwidth]{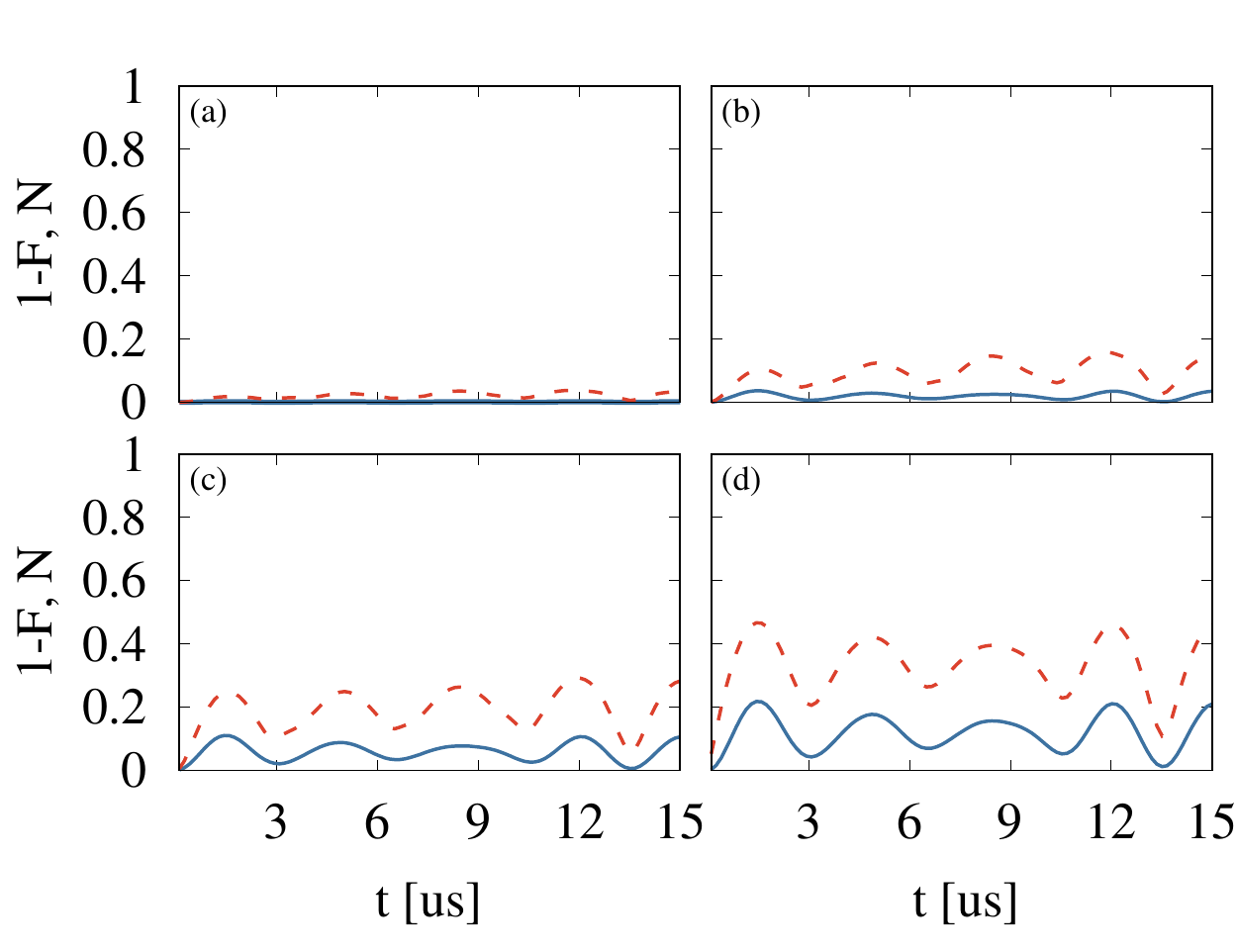}
		\caption{Evolution of qubit-environment Negativity (red dashed lines) and one-minus-Fidelity
			between conditional
			environmental states (blue solid lines) 
			for a qubit defined on $m=0$ and $m=1$ spin states
			and five environmental spins at random locations
			as a function of time for $B_z=0.2$ T and different initial polarizations 
			of the environment: (a) $p=0.1$, (b) $p=0.4$, (c) $p=0.7$, (d) $p=1$.  
		}\label{fig2}
	\end{figure}
	
	We conjecture that the degree of how different the two conditional density
	matrices are, is proportional to the amount of entanglement generated
	throughout the evolution.
	To quantify this difference we will use the Fidelity between $\hat{R}_{nn}(t)$ and $\hat{R}_{11}(t)$, which yields a number between zero and one, one meaning that
	the states are the same and zero that they have orthogonal supports.
	The definition of Fidelity 
	for two arbitrary density matrices (of the same dimensionality) $\hat{R}_{nn}$ and $\hat{R}_{11}$
	is
	\begin{equation}
	\label{f}
	F(\hat{R}_{nn},\hat{R}_{11})=\left[
	\tr
	\left(\sqrt{\sqrt{\hat{R}_{nn}}\hat{R}_{11}\sqrt{\hat{R}_{nn}}}\right)\right]^2.
	\end{equation}
	
	\section{Results \label{sec5}} 
	
	In the following we compare the evolution of Negativity
	between one of the two chosen qubits and the environment and 
	one minus the Fidelity between the conditional 
	states of the environment,
	$1-F(\hat{R}_{nn}(t),\hat{R}_{11}(t))$,
	where $n=-1,0$ is specified by the choice of qubit.
	As the aim here is to study exemplary evolutions of the type as can be found in NV-center
	qubits interacting with a nuclear environment, we use the same randomly chosen realization
	of the spin environment in all plots. They correspond to an environment composed of five
	$^{13}$C isotopes (nuclear spin $1/2$) for which their randomly generated spacial arrangement
	determines the coupling constants (\ref{a}).
	
		\begin{figure}[tb]
		\includegraphics[width=1\columnwidth]{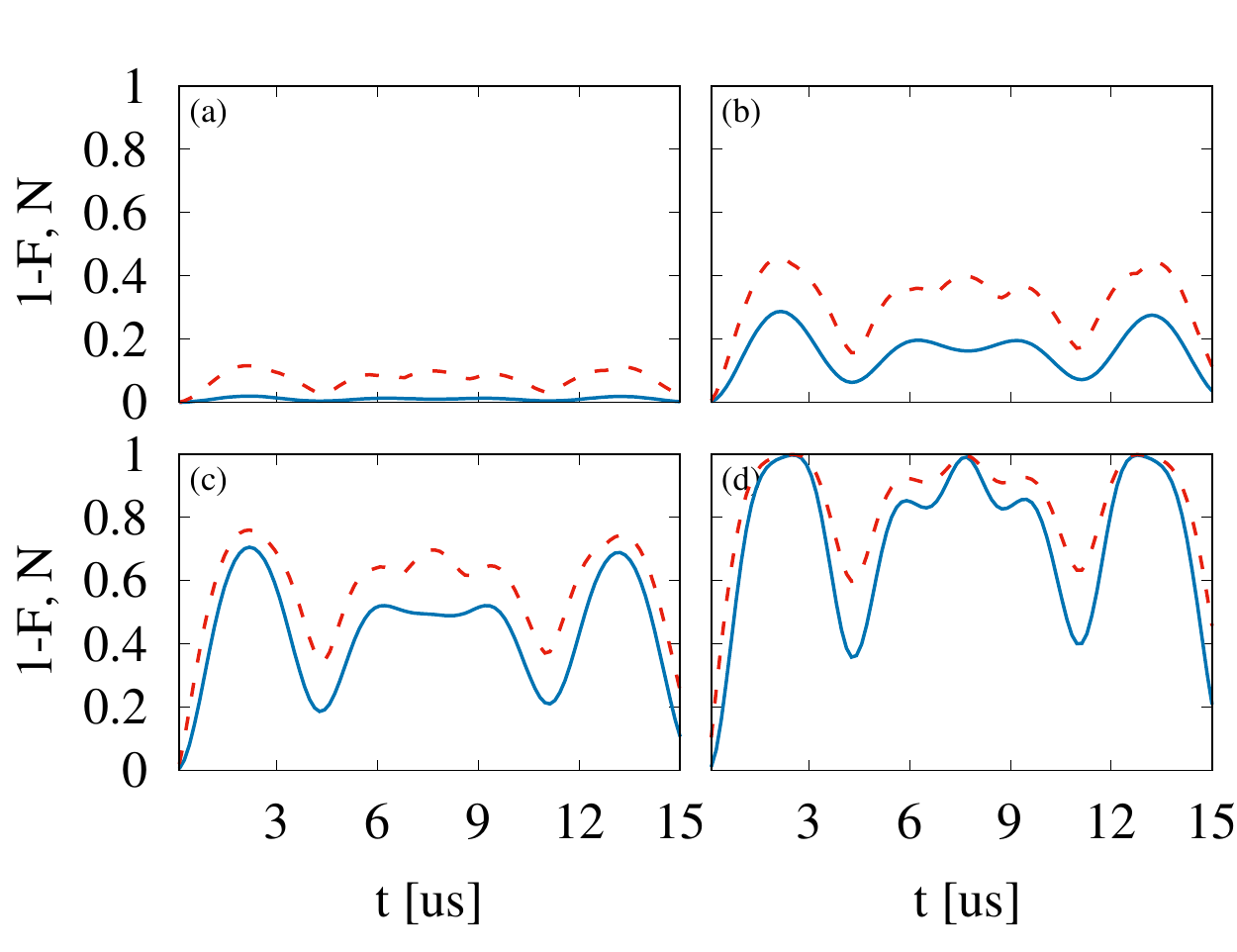}
		\caption{Evolution of qubit-environment Negativity (red dashed lines) and one-minus-Fidelity
			between conditional
			environmental states (blue solid lines) 
			for a qubit defined on $m=-1$ and $m=1$ spin states
			and five environmental spins at random locations
			as a function of time for zero magnetic field and different initial polarizations 
			of the environment: (a) $p=0.1$, (b) $p=0.4$, (c) $p=0.7$, (d) $p=1$.  
		}\label{fig3}
	\end{figure}

	\begin{figure}[tb]
		\includegraphics[width=1\columnwidth]{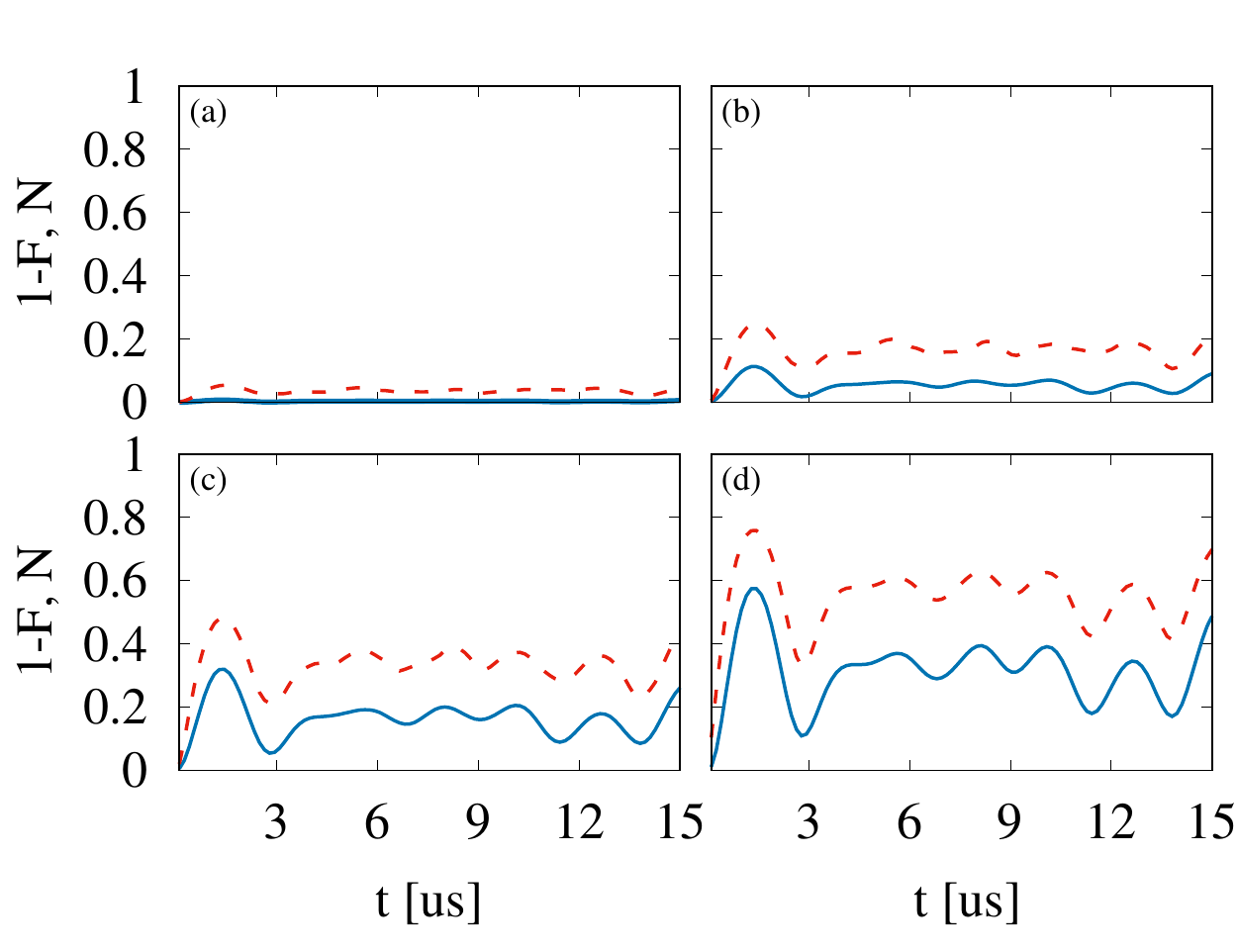}
		\caption{Evolution of qubit-environment Negativity (red dashed lines) and one-minus-Fidelity
			between conditional
			environmental states (blue solid lines) 
			for a qubit defined on $m=-1$ and $m=1$ spin states
			and five environmental spins at random locations
			as a function of time for $B_z=0.2$ T and different initial polarizations 
			of the environment: (a) $p=0.1$, (b) $p=0.4$, (c) $p=0.7$, (d) $p=1$.  
		}\label{fig4}
	\end{figure}
	
	The evolution of Negativity between the $m=0,1$ qubit
	and an environment is plotted
	for $B_z=0$ and $B_z=0.2$ T in Figs \ref{fig1} and \ref{fig2}, respectively,
	using dashed red lines.
	For the $m=-1,1$ qubit analogous plots are found in Fig.~\ref{fig3}
	for $B_z=0$ and in Fig.~\ref{fig4} for $B_z=0.2$ T.
	Analogously, the evolution of the one-minus-Fidelity between the conditional 
	states of the environment, $1-F(\hat{R}_{nn}(t),\hat{R}_{11}(t))$, is plotted in the same
	figures using solid blue lines, with $n=0$ for Figs \ref{fig1} and \ref{fig2}
	and with $n=-1$ for Figs \ref{fig3} and \ref{fig4}.
	The panels (a), (b), (c), and (d) in all plots correspond to growing initial
	polarization of the environment, with the most mixed environment
	(corresponding to $p=1$, so not maximally mixed) in panels (a)
	and fully polarized environments in panels (d). 
	All of the figures contain results for an initial equal superposition qubit state
	(the initial phase between the components of this superposition is irrelevant).
	
	As should be expected \cite{roszak15a}, for a completely mixed environment, $p=0$, qubit decoherence is not accompanied 
	by the generation of entanglement regardless of the type of interaction with environment, since the initial density matrix of the environment
	is proportional to unity and commutes with any possible environmental evolution operators
	\cite{roszak15a}. 
	This does not mean that the qubit does not experience decoherence and, in fact, the qubit 
	becomes dephased during the evolution in all four of the studied situations with not
	polarized initial states of the environment.
	
	For partially and fully polarized initial environmental states, generation
	of entanglement is observed regardless of the variant of the Hamiltonian under study.
	This has been predicted for the $m=0,1$ qubit when $B_z\neq 0$, which has been used
	to exemplify the scheme for detection of qubit-environment entanglement via operations
	only on the qubit subsystem \cite{roszak19}. The procedure described there could also be
	used to predict the generation of qubit-environment entanglement for the $m=-1,1$ qubit and 
	$B_z\neq 0$. This is because the condition for the procedure described in Ref.~\cite{roszak19}
	to be able to detect qubit-environment entanglement is for the evolution operators
	on the environment conditional on the pointer state of the qubit (\ref{w}) not to commute,
	so 
	\begin{equation}
	\left[\hat{w}_n(t),\hat{w}_1(t)\right]\neq 0,
	\end{equation} 
	with $n=-1,0$ depending on the choice of qubit.
	This condition is met for $B_z\neq 0$, but not for $B_z =0$ when 
	$\hat{w}_{-1}(t)=\hat{w}_1^{\dagger}(t)$ and $\hat{w}_{0}(t)=\unit$.
	
\begin{figure}[tb]
		\includegraphics[width=1\columnwidth]{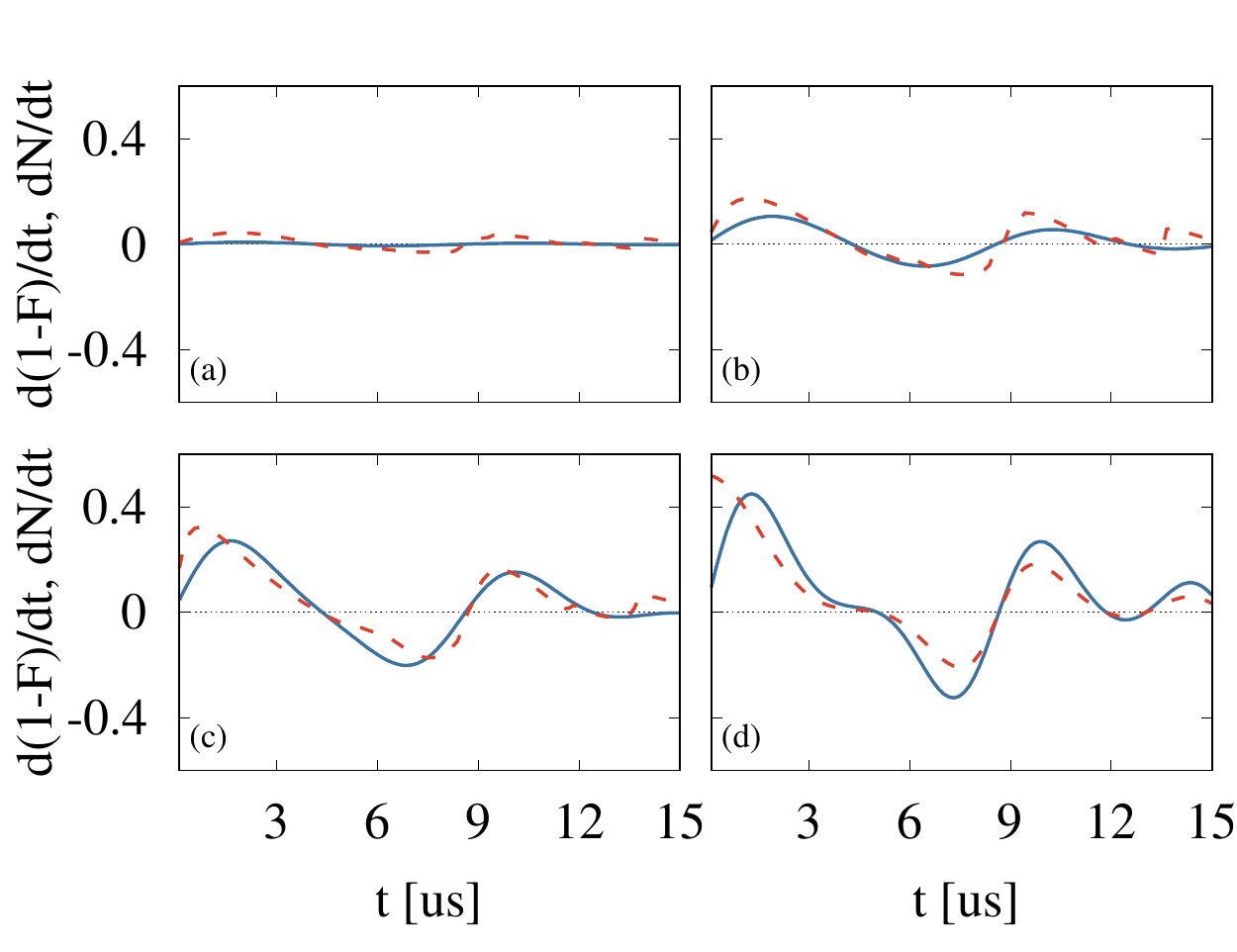}
		\caption{Time-derivative of the evolution of qubit-environment Negativity (red dashed lines) and one-minus-Fidelity
			between conditional
			environmental states (blue solid lines) 
			for a qubit defined on $m=0$ and $m=1$ spin states
			and five environmental spins at random locations
			as a function of time for zero magnetic field and different initial polarizations 
			of the environment: (a) $p=0.1$, (b) $p=0.4$, (c) $p=0.7$, (d) $p=1$. 
			The figure corresponds to the evolutions in Fig.~(\ref{fig1}).
		}\label{fig5}
	\end{figure}

	More interestingly, the evolution of the quantity $1-F(\hat{R}_{nn}(t),\hat{R}_{11}(t))$
	which determines how different the two conditional states of the environment are at time $t$,
	resembles the evolution of the Negativity very closely. In fact,
	$1-F(\hat{R}_{nn}(t),\hat{R}_{11}(t))$ grows when Negativity grows, decreases when Negativity
	decreases, and remains constant when Negativity remains constant. To exemplify this, we plot
	the time-derivatives of both Negativity and one-minus-Fidelity corresponding to the evolutions
	in Fig.~\ref{fig1} in Fig.~\ref{fig5}.
	It can be seen that the derivatives of both quantities are positive, negative, and equal to zero
	at the same segments or points of time.
	
	Since this is the case in all four situations studied, which although they correspond to 
	one physical scenario differ quite extensively, containing an asymmetric system-environment
	coupling (the $m=0,1$ qubit) with ($B_z =0$) and without ($B_z\neq 0$) commuting environmental
	and interaction parts of the Hamiltonian, as well as a coupling which is not asymmetric 
	(the $m=-1,1$ qubit) again in two variants as pertains the commutation of parts of the Hamiltonian,
	it is reasonable to assume that the close resemblance of the Negativity and one-minus-Fidelity
	evolutions is not accidental. 
	
	\section{Conclusion \label{sec6}}
	
	We have studied four variations of an NV-center spin qubit interacting with an 
	environment of a few nuclear spins, which in all cases leads to 
	pure dephasing of the qubit. The variations are obtained by the choice of qubit 
	under study (we chose two out of three possible qubits) which yields different
	effective interaction Hamiltonians and by the application of the magnetic field or 
	lack thereof. The latter facilitates the transition between commuting and non-commuting 
	conditional evolution operators of the environment and is important from the point of
	view of detecting this type of entanglement.
	
	We have compared the time-evolution of the amount of entanglement between the qubit and
	the environment with the time-evolution of one minus the Fidelity of the state of the environment
	at time $t$ on the qubit pointer state. In all studied situations the evolution of one-minus-Fidelity
	resembled the evolution of Negativity very closely, to the extent that both quantities 
	were growing and decreasing in the same time-segments. 
	
	We conjecture that the amount of entanglement
	with the environment generated
	during any evolution that leads to pure dephasing of the qubit for an initial product state
	of a pure qubit and environment is proportional to one-minus-Fidelity between the 
	states of the environment conditional on the qubit pointer states.
	This would mean that the amount of entanglement generated between the qubit and the
	environment is proportional to the trace that the joint evolution leaves on the environment.
	Hence, although it is not possible to distinguish between entangling and nonentangling
	evolutions by studying the level of qubit dephasing, it is not only possible to distinguish them
	by detecting the difference in environmental evolution linked to the different pointer states
	of the qubit,
	but it may also be possible to quantify the amount of entanglement in the qubit-environment
	system by studying the magnitude of this difference.
	We have shown this to be the case in quantitatively different situations which can be 
	realized in NV-center spin qubits.
	The advantage is that contrary to other measures of mixed-state entanglement, here
	we have a natural physical interpretation,
	which in fact is the same as for the pure state entanglement
	in pure dephasing scenarios (namely, how much the two conditional states of the environment differ
	from one another).

	This work was supported by funds of the Polish National Science Center (NCN), Grant no.~2015/19/B/ST3/03152.

\end{document}